\input{aipcheck}

\documentclass[
    ,final            % use final for the camera ready runs
  ]
  {aipproc}

\usepackage{amsmath}
\usepackage{graphicx}
\usepackage[varg]{txfonts} % Times fonts
\usepackage[latin1]{inputenc}
\usepackage{bm}

\layoutstyle{8x11single}

\begin{document}

\title{Hadron Spectroscopy at CLEO}


\author{Amiran Tomaradze (for the CLEO Collaboration)}{
  address={Department of Physics and Astronomy, 
        Northwestern University, Evanston, IL, 60208, USA}
}

\begin{abstract}
New measurements of the masses and
decay branching fractions of charmonium and bottomonium 
states using the data collected by the CLEO
detector are presented.  These include CLEO identification
of the singlet states  $\eta_c'(2S)$, $h_{c}(1P)$, and $\eta_{b}(1S)$.
Comparison with other experimental measurements and theoretical 
models is also presented.
\end{abstract}

\maketitle

\section{Introduction}

The \textbf{QCD interaction} 
 can be studied in light quark ($u,d,s$) hadrons as well as 
heavy quark ($c,b$) hadrons. 
In contrast to light quarks, heavy quark states are narrow
and do not mix with the states of other quarks. 
This is illustrated in Fig.~1(left) for charmonium.
Also, the effective coupling constant and relativistic problems are far 
more tractable. Thus, the  
spectra of charmonium and bottomonium are easier to characterize and study.

\section{CLEO Data for Charmonium and Bottomonium Spectroscopy}

The world's largest pre--BESIII sample of 26 million $\psi(2S)$ 
comes from CLEO. These $\psi(2S)$ data have been used to study 
the spectroscopy of 
$\chi_{cJ}(^{3}P_{J})$ and $h_{c}(^{1}P_{1})$.
Using  $\pi\pi$ tag in the decay 
 $\psi(2S)\to \pi^{+}\pi^{-}J/\psi$ ($B$=35\%), the spectroscopy of
$J/\psi$ is also studied.

CLEO collected a sample of 21 million $\Upsilon(1S)$, 9 million 
$\Upsilon(2S)$, and 6 million $\Upsilon(3S)$. Besides bottomonium 
spectroscopy, the $\Upsilon$ data are used for charmonium 
spectroscopy using two-photon fusion reactions.

My talk contains two parts: (a)
 CLEO measurements of the masses of charmonium and
bottomonium singlet states $\eta_c'(2S)$, $h_{c}(1P)$,
and $\eta_{b}(1S)$, and  their implications for the $q\bar{q}$
 hyperfine interaction; (b) CLEO measurements for the decay 
branching fractions of charmonium and bottomonium states.

\begin{figure}[!tb]
\includegraphics[width=3.in]{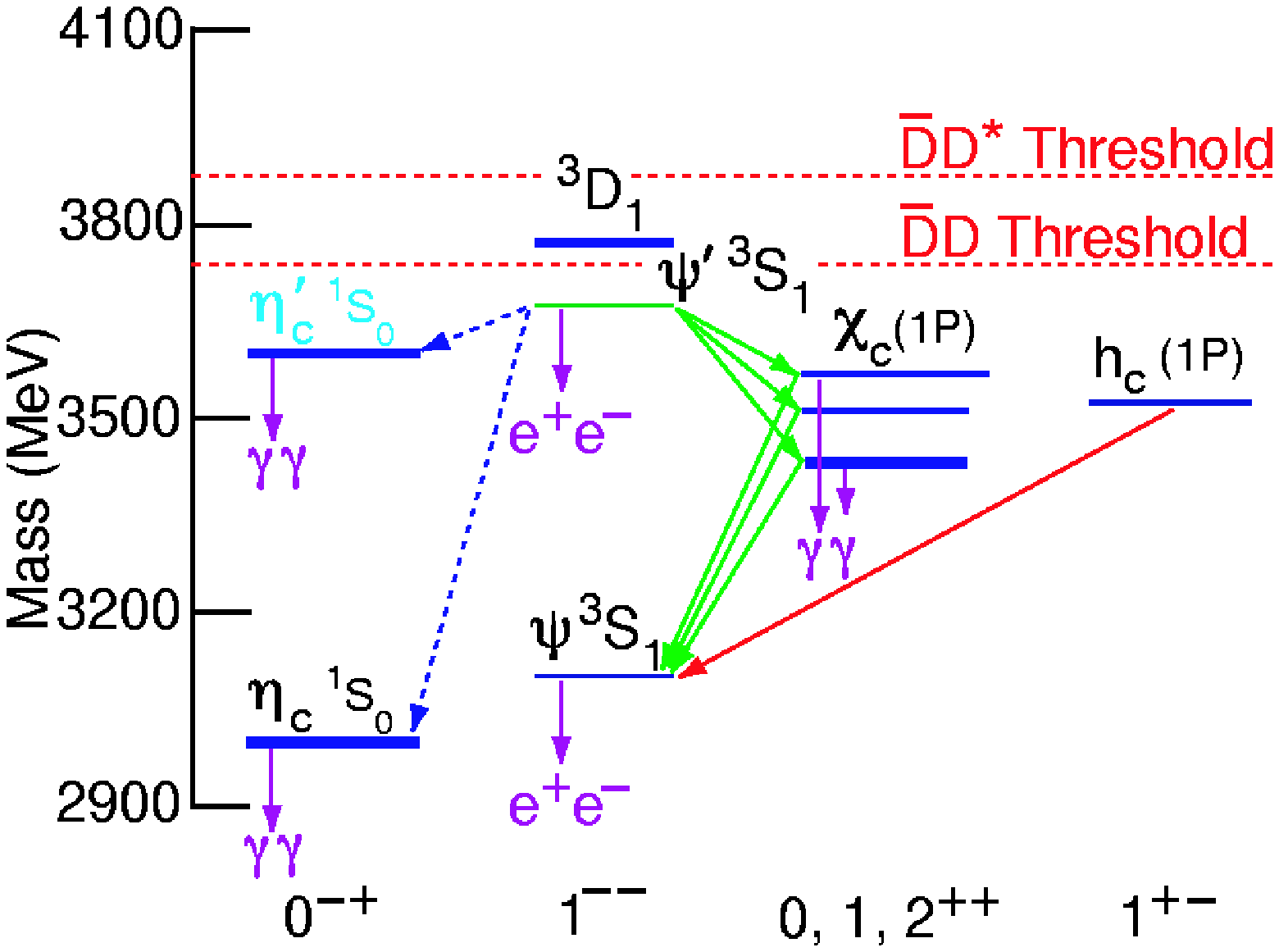}
\includegraphics[width=2.2in]{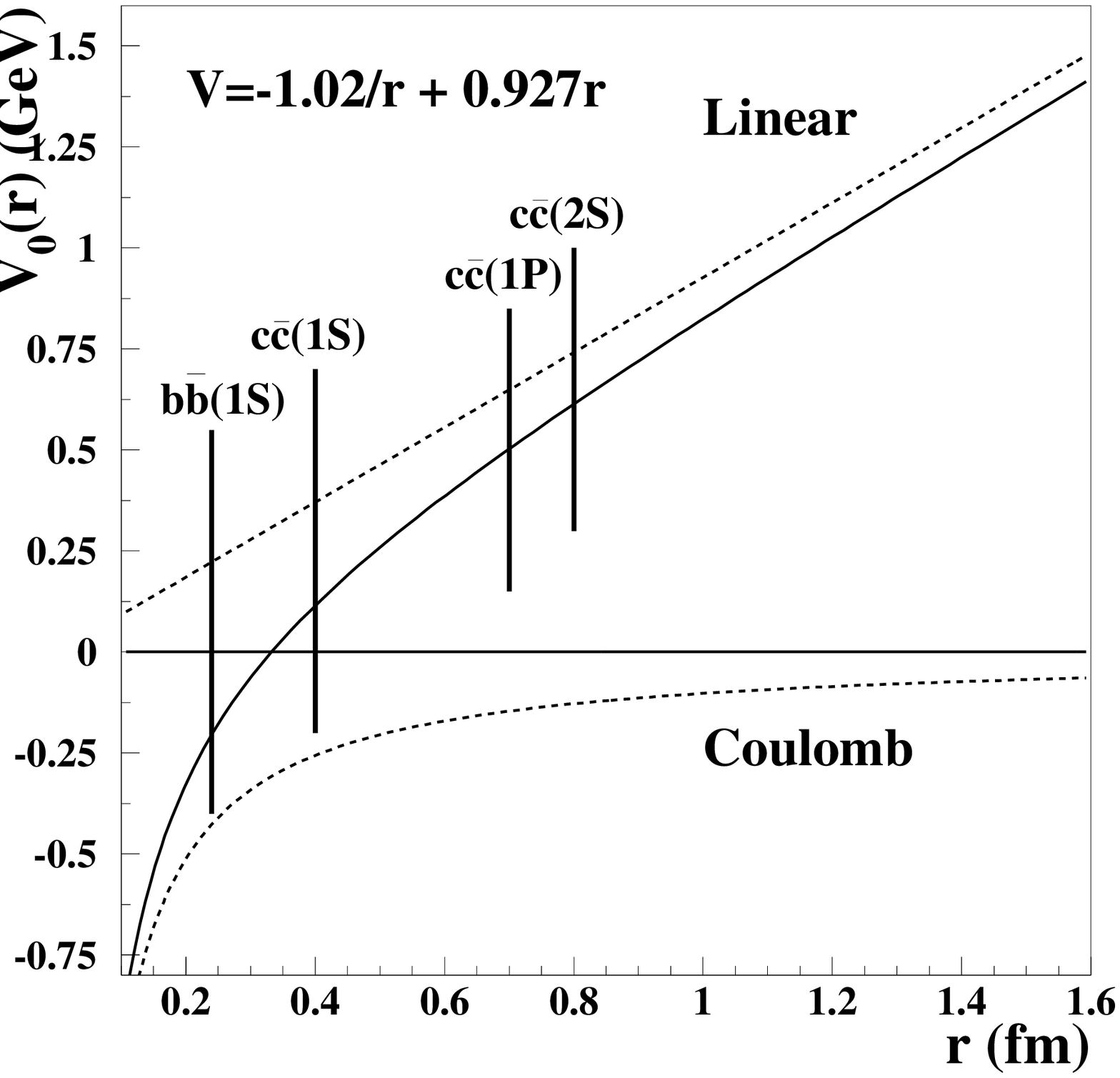}
\caption{(left) Spectra of the states of Charmonium.
 (right) Schematic of the QCD $q\bar{q}$ potential (solid line), 
and its Coulombic and confinement parts (dotted lines).
The vertical lines show the approximate location of the 
$\left|c\bar{c}\right>$ charmonium and $\left|b\bar{b}\right>$ 
bottomonium bound states. }
\end{figure}

\section{The $q\bar{q}$ Hyperfine Interaction}

In the quark model the hyperfine spin--spin interaction determines  
the \textbf{ground-state masses} of the hadrons. The
 mass of a pseudoscalar or vector $q\bar{q}$ meson is
$$M(q_1\bar{q}_2) = m_1(q_1) + m_2(q_2) + A \left[ \frac{\vec{s}_1\cdot\vec{s}_2}{m_1m_2} \right].$$
The $\vec{s_1}\cdot\vec{s_2}$ spin--spin, or 
\textbf{hyperfine interaction} gives
 rise to the hyperfine, or spin-singlet/spin-triplet splitting
in quarkonium spectrum,
$$\Delta M_{hf}(nL) \equiv  M(n^3L_J) - M(n^1L_{J=L}).$$

The hyperfine interaction is not well understood because 
until recently there were not enough experimental data to provide
the required constraints for the theory.
For thirty years after the discovery of $J/\psi$, 
the only hyperfine splitting measured in a hidden flavor meson was
$\bm{\Delta M_{hf}(1S)_{c\bar{c}}\equiv M(J/\psi)-M(\eta_c) = 116.4\pm1.2~\mathrm{MeV}}$~\cite{pdg}.
No other singlet states, 
$\bm{\eta_c'(2^1S_0)_{c\bar{c}},~h_c(1^1P_1)_{c\bar{c}},~\mathrm{or}~\eta_b(1^1S_0)_{b\bar{b}}}$
 were identified, and none of the 
important questions about the hyperfine interaction could be answered. 
\textbf{This has changed in the last few years.}

\subsection{$\eta_c'(2S)$, Hyperfine Splitting in a Radial Excitation}

In 2002, Belle claimed identification of $\eta_c'$ in the decay of 45 
million $B$ mesons, $B\to K(K_SK\pi)$ and reported 
$M(\eta_c')=3654\pm10$ MeV, which would correspond to 
$\Delta M_{hf}(2S)=32\pm10$ MeV~\cite{belle-etacpb}, a factor \textbf{two smaller} than 
expected and a factor \textbf{four} smaller than $\Delta M_{hf}(1S)$.
It became important to confirm this result.

There are two important ways $2S$ states differ from $1S$ states.  
$1S$ states, with $r\approx0.4~\mathrm{f}$, lie in the Coulombic 
region ($\sim1/r$) of the $q\bar{q}$ potential, $V=A/r+Br$, 
whereas the $2S$ states, with $r\approx0.8~\mathrm{f}$, lie in 
the confinement part ($\sim{r}$) of the potential (see Fig.~1, right).  
The spin--spin potential in the two regions could be different.  
The second difference is that the $2S$ states, particularly $\psi(2S)$, 
lie close to the $D\bar{D}$ breakup threshold at 3730 MeV, and can 
be expected to mix with the continuum as well as 
higher $1^{--}$ states.  All in all, it is important to nail 
down $\eta_c'$ experimentally, and measure its mass accurately.

This was successfully done by  CLEO~\cite{cleo-etacp} and 
BaBar~\cite{babar-etacp} in 2004 by observing $\eta_c'$ in two--photon 
fusion, $\gamma\gamma\to\eta_c'\to K_SK\pi$.  
The two observations are shown in Fig.~2.  The average of all 
measurements is $M(\eta_c')=3637\pm4$ MeV~\cite{pdg}, which leads 
to $\bm{\Delta M_{hf}(2S)=49\pm4}$ MeV, which is almost 
a factor \textbf{2.5 smaller} than $\Delta M_{hf}(1S)$.  
Explaining this large difference is a challenge to the theory.  
The challenge for the experimentalists lies in completing the 
spectroscopy of $\eta_c'$, now that its mass is known.
In particular, it is important to measure its width.

\begin{figure}[!tb]
%\begin{center}
\includegraphics[width=2.5in]{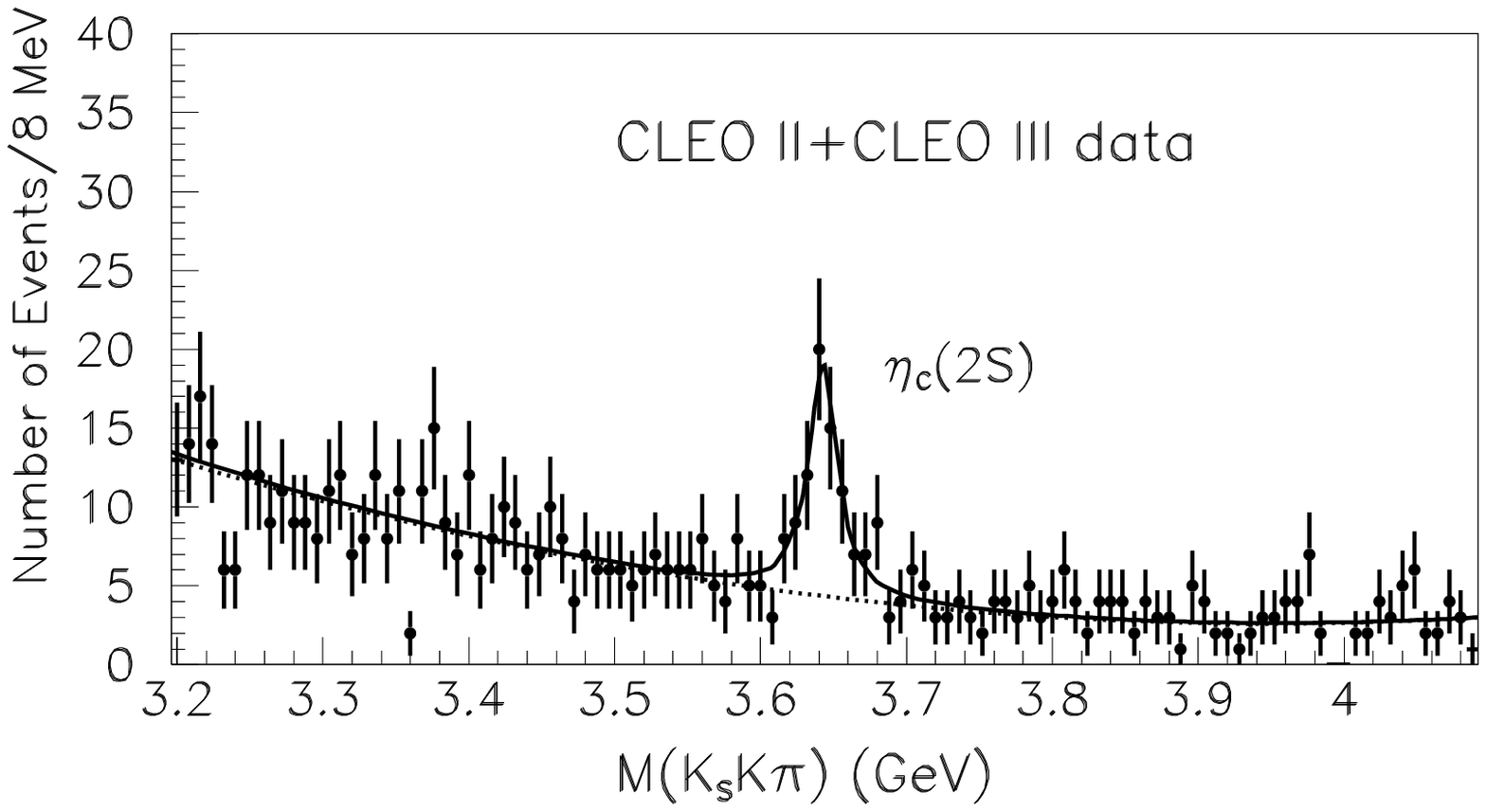}
\includegraphics[width=2.4in]{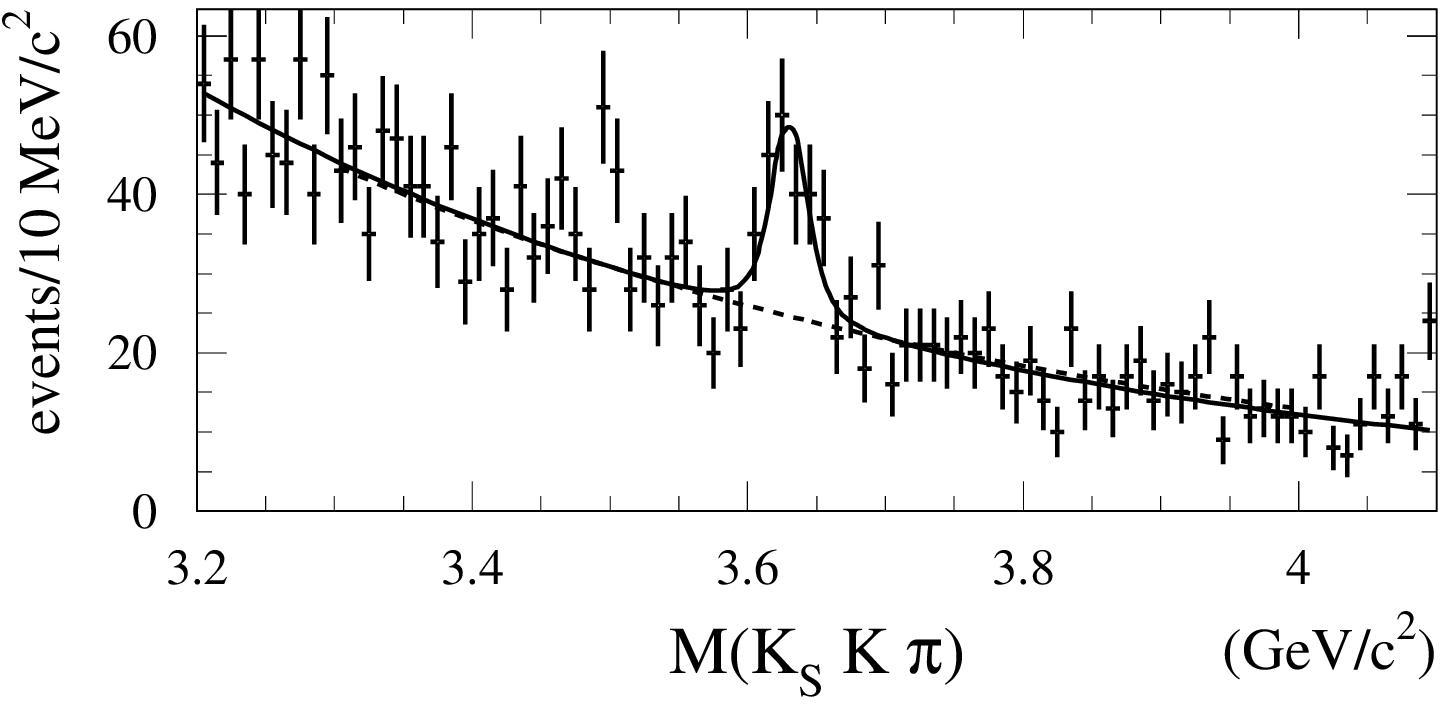}
%\end{center}
\caption{The invariant mass $M(K_SK\pi)$ spectra from two--photon fusion measurements by CLEO (left) and BaBar (right).  The $\eta_c(2S)$ peak is prominent in both spectra.}
\end{figure}

\subsection{$h_c(1^1P_1)$, Hyperfine Interaction in $P$--wave}

In this case, we have a very simple, and provocative theoretical expectation, namely
\begin{equation}
\Delta M_{hf}(1P) \equiv M(^3P) - M(^1P) = 0.
\end{equation}
This arises from the fact that a non-relativistic reduction of the Bethe-Salpeter equation makes the hyperfine interaction a \textbf{contact interaction}.  Since only S--wave states have finite wave function at the origin,
\begin{equation}
\Delta M_{hf}(L\ne0)=0.
\end{equation}
We can test this prediction in charmonium by
\begin{itemize}
\item identifying the singlet--P state $h_c(1^1P_1)$, and
\item by estimating $M(^3P)$, given the masses of the triplet--P states $\chi_{0,1,2}~(^3P_{0,1,2})$.
\end{itemize}

\begin{figure}
\includegraphics[width=2.0in]{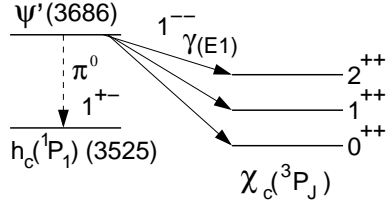}
\caption{Comparing allowed E1 transitions from $\psi'(^3S_1)$ to $\chi_{cJ}(^3P_J)$ states of charmonium with the isospin forbidden $\pi^0$ transition to the singlet P--state $h_c(^1P_1)$.}
\end{figure}

The experimental identification of $h_c(1^1P_1)$ is even more difficult than that of $\eta_c'$.  The centroid of the $^3P_J$ states is at 
$3525.30\pm0.04$~MeV~\cite{pdg}. If Eq.~1 is true, $M(h_c)\approx3525$~MeV, i.e., $\sim160$~MeV below the $\psi(2S)$ state from which it must be fed.  Unfortunately, populating $h_c$ has several problems.
\begin{itemize}
\item The radiative transition $\psi(2S)(1^{--})\to\gamma h_c(1^{+-})$ is forbidden by \textbf{charge conjugation} invariance.  
\item The only other alternative is to populate $h_c$ in the reaction 
$\psi(2S)\to\pi^0h_c$.
But that is not easy, because a $\pi^0$ transition ($M(\pi^0)=139$~MeV) has very little phase space, and further, the reaction is forbidden by \textbf{isospin conservation}. Nevertheless, this is the only possible way of populating $h_c$, and we at CLEO had to valiantly go for it.
An illustration of the allowed E1 transitions from $\psi(2S)(^3S_1)$ to 
$\chi_{cJ}(^3P_J)$ states and the isospin forbidden $\pi^0$ transition to the singlet P--state $h_c(^1P_1)$ is shown in Fig.~3.
\end{itemize}

In 2005, we at CLEO made the first firm identification (significance$>6\sigma$) of $h_c$ in the reaction 
$$\psi(2S)\to\pi^0h_c,~~h_c\to\gamma\eta_c,$$ 
%which is illustrated in Fig.~4.  
in an analysis of 3.08~million $\psi(2S)$ decays~\cite{cleo-hc}.

\begin{figure}
\includegraphics[width=2.5in]{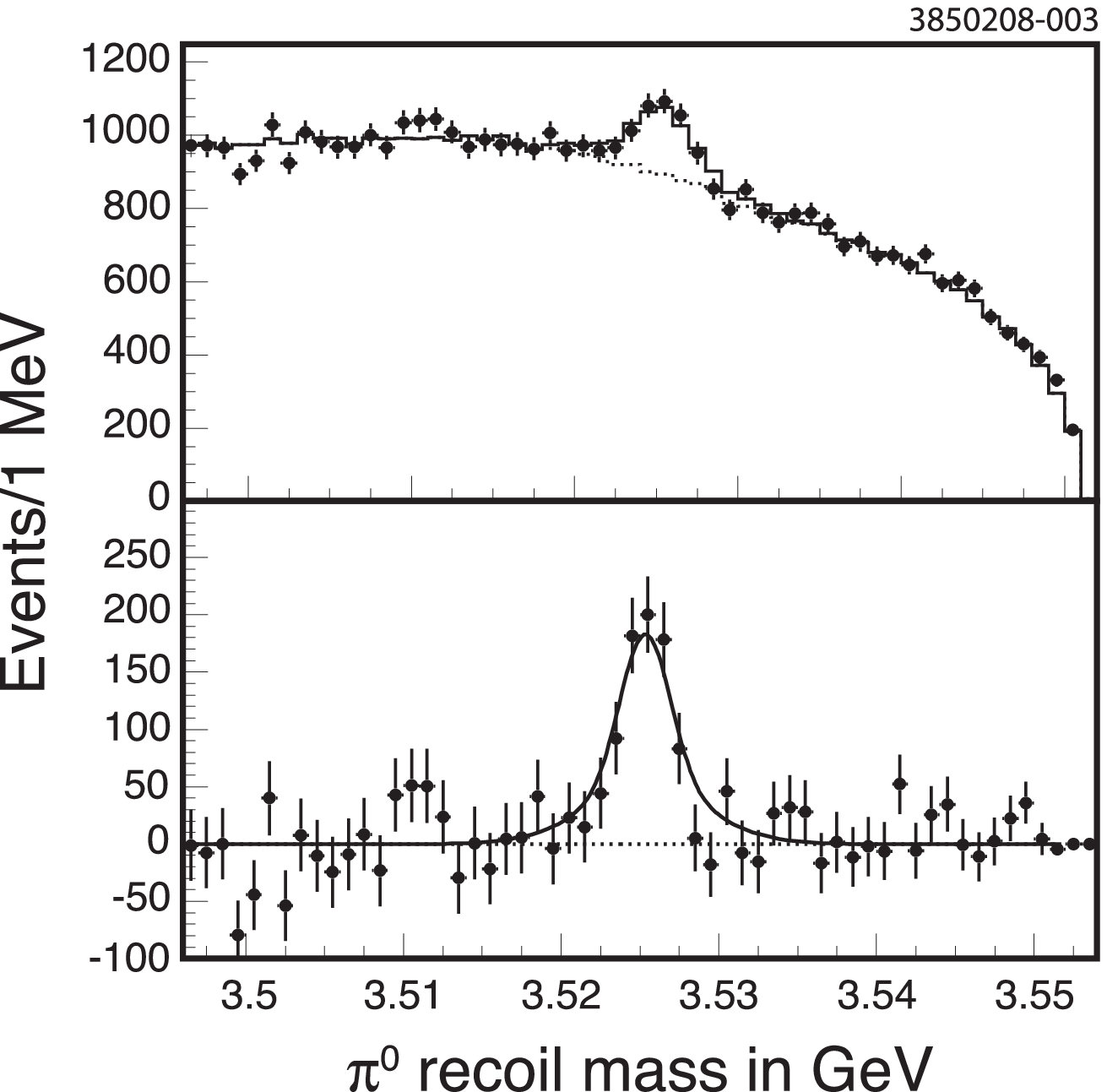}
\includegraphics[width=2.5in]{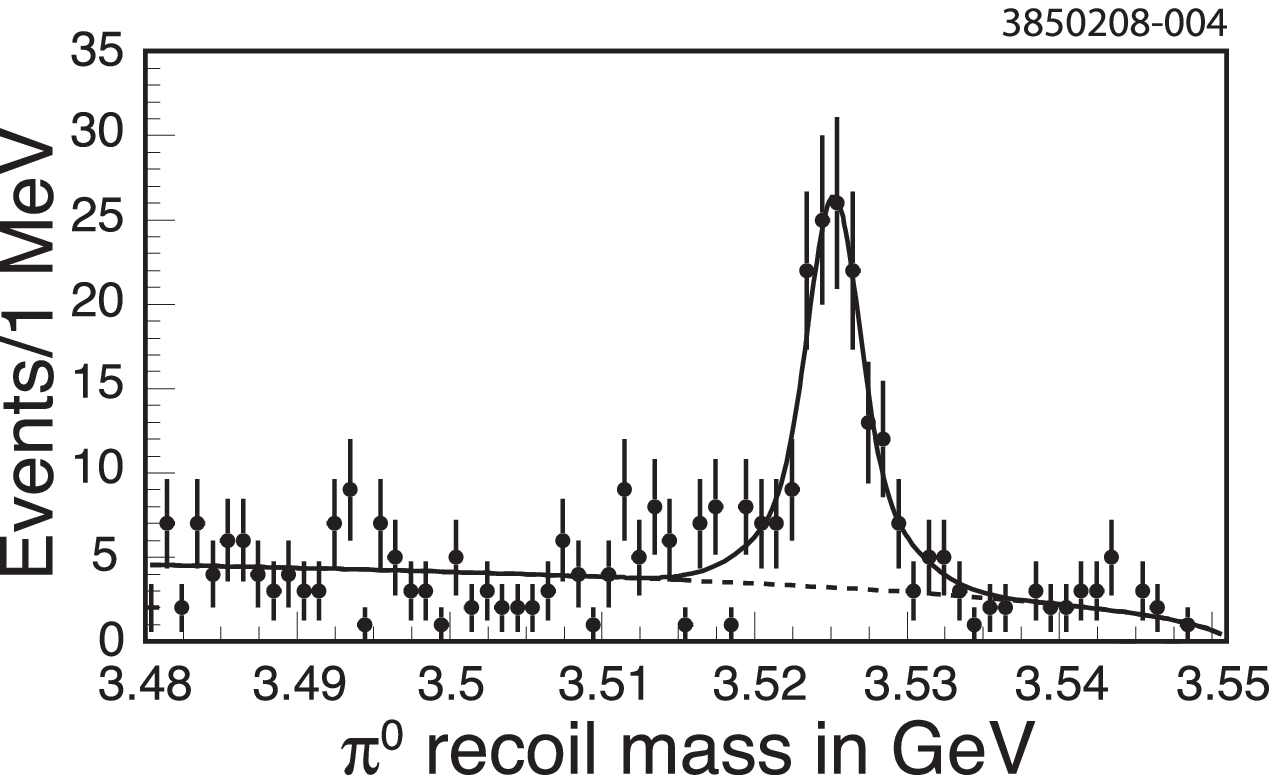}
\caption{The recoil mass of $\pi^0$ in the decay $\psi(2S)\to\pi^0h_c$.  (left) Full and background subtracted spectra for inclusive analysis.  (right) Spectrum of exclusive analysis.}
\end{figure}

In 2008, we repeated our measurement with 8 times larger luminosity, and  24.5 million $\psi(2S)$~\cite{cleo-hc-new}.  As before, data were analyzed in two ways.  In the inclusive analysis, the photon energy, $E_\gamma$, was loosely constrained, but the decay products of $\eta_c$ were not identified. In the exclusive analysis, instead of constraining $E_\gamma$ fifteen hadronic decay channels of $\eta_c$ were measured. 
As shown in Fig.~4, $h_c$ was observed with significance~~$>13\sigma$.
The total number of events was $N(h_c)=1146\pm118$ from inclusive 
analysis, and $N(h_c)=136\pm14$ from exclusive analysis. 
The results from inclusive and exclusive analyses were consistent.
The precision results were 
\begin{align}
M(h_c) & =3525.28\pm0.19\pm0.12~\mathrm{MeV},\\
\nonumber\mathcal{B}_1\times\mathcal{B}_2 & =(4.19\pm0.32\pm0.45)\times10^{-4}.
\end{align}
% $\bm{h_c(^1P_1)}$ \textbf{is now firmly established.} 
Thus, $h_c(^1P_1)$ \textbf{is now firmly established.}

If it is assumed that $M(^3P)$ is identical to the centroid of the triplet--P states, $\left<M(^3P_J)\right>=[5M(\chi_{c2})+3M(\chi_{c1})+M(\chi_{c0})]/9=3525.30\pm0.04$~MeV, then the above $M(h_c)$ leads to the hyperfine splitting,
\begin{equation}
\Delta M_{hf}(1P)_{c\bar{c}}=\left<M(^3P_J)\right>-M(^1P_1)=0.02\pm0.23~\mathrm{MeV,}
\end{equation}
but, $\left<M(^3P_J)\right>_{0,1,2} \ne M(^3P)$!

The centroid  $\left<M(^3P_J)\right>$ is a good measure of $M(^3P)$ only if the spin--orbit splitting between the states $^3P_2$, $^3P_1$, and $^3P_0$ is perturbatively small.  It is obviously not so.
The splitting, $M(^3P_2)-M(^3P_0)=142$~MeV, is not small.
Further, the perturbative prediction is that 
\begin{align}
M(^3P_1)-M(^3P_0)& =\frac{5}{2}\left[M(^3P_2)-M(^3P_1)\right]=114~\mathrm{MeV},
%\nonumber   & =114~\mathrm{MeV},
\end{align}
while the experimental value is
\begin{equation}
M(^3P_1)-M(^3P_0)=96\pm1~\mathrm{MeV}.
\end{equation}
This is a 18 MeV difference! So we are obviously not in the 
perturbative regime.

This leads to serious questions.
\begin{itemize}
\item What mysterious cancellations are responsible for the wrong estimate of $M(^3P)$ giving the expected answer that
$$\Delta M_{hf}(1P)=0.$$
\item Or, is it possible that the expectation is wrong?  Is it possible that the hyperfine interaction is not entirely a \textbf{contact interaction}?  
\item Potential model calculations are not of much help because they smear the potential at the origin in order to be able to do a Schr\"odinger equation calculation.
\item Can Lattice help? So far we have no lattice predictions with
sufficient precision.
\end{itemize}

\subsection{$\eta_b(^1S_0)$, Hyperfine Interaction Between $b$--Quarks}

The $b\bar{b}$ bottomonium system is, in principle, the best one
 to study the fundamental aspects of the hyperfine interaction 
between quarks.   Unfortunately, until last year we had no knowledge 
of the hyperfine interaction between $b$--quarks.  
The spin--triplet $\Upsilon(1^3S_1)$ state of bottomonium was 
discovered in 1977, but its partner, the 
spin--singlet $\eta_b(1^1S_0)$ ground state of bottomonium, was not 
identified for thirty years, mainly because of  
the difficulty in observing weak M1 radiative 
transitions.  There were many pQCD based theoretical predictions  
which varied all over the map, with $\Delta M_{hf}(1S)_b=35-100$~MeV, 
and $\mathcal{B}(\Upsilon(3S)\to\gamma\eta_b)=(0.05-25)\times10^{-4}$.  

This has changed now.  The $\eta_b(1^1S_0)$ ground state of the
$\left|b\bar{b}\right>$ Upsilon family \textbf{has been finally identified!}

In July 2008, BaBar announced the identification of $\eta_b$~\cite{ups-babar}.  They analyzed the inclusive photon spectrum of
\begin{equation}
\Upsilon(3S)\to\gamma\eta_b(1S)
\end{equation}
in their data for \textbf{120 million} $\Upsilon(3S)$ (28~fb$^{-1}~e^+e^-$). 
 BaBar's success owed to their very large data set and a clever way of 
reducing the continuum background, a cut on the so--called thrust angle, 
the angle between the signal photon and the thrust vector of the 
rest of the event,  $|\cos\theta_{Thrust}|<0.7$. BaBar's results were:
\begin{gather}
M(\eta_b)=9388.9^{+3.1}_{-2.3}\pm2.7~\mathrm{MeV},\\
\nonumber \Delta M_{hf}(1S)_b=71.4^{+3.1}_{-2.3}\pm2.7~\mathrm{MeV}, \\
\nonumber \mathcal{B}(\Upsilon(3S)\to\gamma\eta_b) = (4.8\pm0.5\pm0.6)\times10^{-4}.
\end{gather}

The significance of $\eta_b$ observation was $>$10$\sigma$. 
Recently, BaBar has also reported a 3.0$\sigma$ identification of
$\eta_b$ in $\Upsilon(2S)\to\gamma\eta_b$~\cite{ups-babar}.

Any important discovery requires confirmation by 
an \textbf{independent} experiment. At CLEO we 
had data for only \textbf{5.9~million} $\Upsilon(3S)$, i.e., 
about 20~times less than BaBar.    
But we have better photon energy resolution, and we have been able 
to improve on BaBar's analysis technique.  We make three improvements.  
We make very detailed analysis of the large continuum background under 
the resonance photon peaks.  We determine photon peak 
shapes by analyzing background from peaks in background--free radiative 
Bhabhas and in exclusive $\chi_{b1}$ decays.  And we make a joint 
fit of the full data in three bins of $|\cos\theta_T|$, covering 
the full range $|\cos\theta_T|=0-1.0$ (see Fig.~5).  
Monte-Carlo simulations show that the joint fit procedure leads to  
an average increase of the significance of an $\eta_b$ signal by $\sim20\%$
over accepting only events with  $|\cos\theta_{Thrust}|<0.7$.
So, despite our 
poorer statistics, we have succeeded in confirming BaBar's discovery with
significance level $\sim$4$\sigma$.  The results have been submitted for 
publication~\cite{ups-cleo-new}. 

Our results are:
\begin{gather}
M(\eta_b)=9391.8\pm6.6\pm2.0~\mathrm{MeV},\\
\nonumber \Delta M_{hf}(1S)_b = 68.5\pm6.6\pm2.0~\mathrm{MeV},\\
\nonumber \mathcal{B}(\Upsilon(3S)\to\gamma\eta_b)=(7.1\pm1.8\pm1.3)\times10^{-4}.
\end{gather}

\begin{figure}
\includegraphics[width=3.0in]{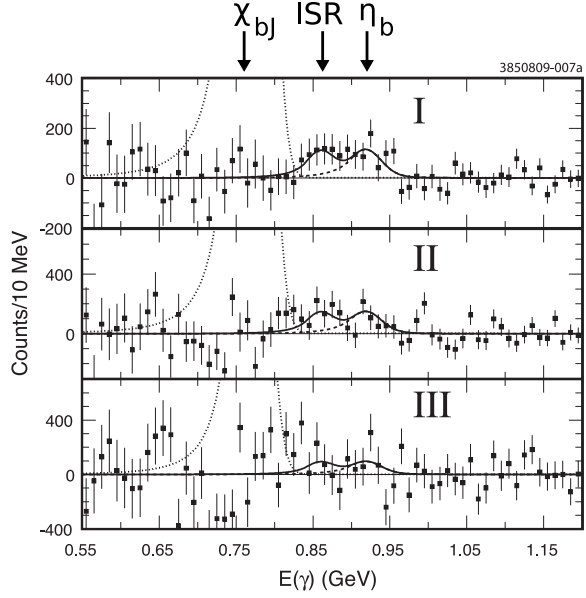}
\caption{Illustrating CLEO results for the identification of $\eta_b$ in a joint fit of data in three bins of the thrust angle,  I:~$|\cos\theta_T|=0-0.3$,  II:~$|\cos\theta_T|=0.3-0.7$,  III:~$|\cos\theta_T|=0.7-1.0$.}
\end{figure}

 The results agree with those of BaBar. The average of our and BaBar's 
results for the hyperfine splitting is
$$\left<\Delta M_{hf}(1S)_b\right> \equiv M(\Upsilon(1S))-M(\eta_b) = 69.4\pm2.8~\mathrm{MeV}.$$
A recent unquenched lattice calculation predicts 
(NRQCD with $u,d,s$ sea quarks) $\Delta M_{hf}(1S)_b=61\pm14$~MeV.  
A quenched lattice calculation (chiral symmetry and $s,c$ sea quarks) 
predicts $\Delta M_{hf}(1S)_b=70\pm5$~MeV.
Thus, as far as the hyperfine splitting for the $\left|b\bar{b}\right>$ 
is concerned, lattice calculations appear to be on the right 
track~\cite{lqcd}.

For more details on $\eta_b$ analysis by CLEO see the talk by
S. Dobbs in the parallel session 7C~\cite{dobbs}.

\subsection{Hyperfine Splittings Measurements}

To summarize, we now have well--measured experimental results for 
several hyperfine splittings, with significant contributions 
from CLEO measurements.\\\\
\begin{tabular}{ll}
$\left|c\bar{c}\right>$ Charmonium: &   $\Delta M_{hf}(1S)=116.4\pm1.2$~MeV, \\[2pt]
 &    $\Delta M_{hf}(2S)=49\pm4$~MeV, \\[2pt]
 &    $\Delta M_{hf}(1P)=0.02\pm0.23$~MeV,  \\[4pt]
$\left|b\bar{b}\right>$ Bottomonium: &   $\Delta M_{hf}(1S)=69.4\pm2.8$~MeV. \\[6pt]
\end{tabular}

In charmonium, we do not have satisfactory understanding of the variation of hyperfine splitting for the S--wave radial states, and for P--wave state.
\begin{itemize}
\item For charmonium, we do not have any unquenched lattice predictions, at present.
\item For bottomonium, lattice predictions are available, and they appear to be on the right track.
\item For neither charmonium nor bottomonium there are any reliable predictions of transitions strength, particularly for forbidden M1 transitions.
\end{itemize}

Much remains to be done.  On the experimental  front it is very important to identify for bottomonium the allowed M1 transition, $\Upsilon(1S)\to\gamma\eta_b(1S)$, and to identify the bottomonium singlet P--state, $h_b(^1P_1)$.  On the theoretical front one would like to see unquenched lattice calculations for charmonium singlets, and, of course, for transition strengths.

\section{Measurements of the decay branching fractions of charmonium and bottomonium states}

\subsection{ Search for Exclusive Decays of $\eta_c'(2S)$}

Recently, CLEO has performed a search for the decay 
$\psi(2S) \rightarrow \gamma \eta_c'(2S)$ in a sample 
of 26 million $\psi(2S)$ events~\cite{cleo-etacp1}. 
Expected $E_{\gamma}=48$ MeV. 
Eleven exclusive decay modes, 
$\eta_c'(2S)\to\mathrm{hadrons},~(\pi,~K,~\eta,~\eta')$ with up 
to 6 particles (charged and neutrals) were reconstructed, but
no signals of $\eta_c'(2S)$ were observed in any of 
the decay modes, or in their sum. 

The product branching fraction upper limits were determined for the individual
modes, and they are at the level of (4--15)$\times 10^{-6}$. 
These upper limits are an order of magnitude smaller than expected by 
assuming that the partial widths for $\eta_c'(2S)$ decays are the same 
as for  $\eta_c(1S)$.

 Thus, so far $K_SK\pi$ is the only decay mode in which
$\eta_c'(2S)$ has been identified.

\subsection{Evidence for Exclusive Decay of $h_c(1P)$ to Multipions}

Now that $h_c$ has been discovered, CLEO has searched 
for hadronic decays of $h_c$ in multipion channels~\cite{hc-je}. 
Of the three decays investigated, only one,  the five pion decay 
$h_c\to 2(\pi^+\pi^-)\pi^0$,
is found to have a statistically significance signal, with 
$B(\psi(2S)\to h_c)\times B(h_c \to 2(\pi^+\pi^-)\pi^0)=
(1.9^{+0.7}_{-0.5})\times10^{-5}$ (see Table I). 
This is $\sim5\%$ of 
$B(\psi(2S)\to h_c)\times B(h_c \to \gamma\eta_c)=(4.19\pm0.32\pm0.45)\times10^{-4}$. \\

\begin{table}
\begin{tabular}{c|c|c|c}
\hline
Mode  & Efficiency (\%) & Yield & 
 $B(\psi(2S)\to h_c)\times B(h_c \to n(\pi^+\pi^-)\pi^0)\times10^5$ \\
\hline
$\pi^+\pi^-\pi^0$ &27.0\% & $1.6^{+6.7}_{-5.9}$ &$ <0.2$ (90\%)  \\
$2(\pi^+\pi^-)\pi^0$ & 18.8\% & $92^{+23}_{-22}$ & $1.88^{+0.48+0.47}_{-0.42-0.16}$~(significance $\sim4\sigma)$\\ 
$3(\pi^+\pi^-)\pi^0$ & 11.5\% & $35\pm26$ & $<2.5$ (90\%) \\
\hline
\end{tabular}
\caption{Results for exclusive decays of $h_c(1P)$ to multipions.}
\end{table}

\subsection{Observation of $J/\psi\to 3\gamma$}

 No $3\gamma$ decay of a meson has been observed before.
 In the lowest order, $3\gamma$ decay   
is a QED process, and the predicted ratio 
$\mathcal{B}(J/\psi\to3\gamma)/\mathcal{B}(J/\psi\to e^+e^-)=5.3\times10^{-4}$, which is independent of charm quark mass and wave function, leads to
$\bm{\mathcal{B}(J/\psi\to3\gamma)=3.2\times10^{-5}}$.
QCD radiative corrections, which are not reliably known, may modify 
the prediction.

To search for $3\gamma$ decay of $J/\psi$,  
CLEO has used a QED background free sample of 
9.6 million $J/\psi$ obtained by $\pi^{+}\pi^{-}$ tagging
in the decay $\psi(2S)\to (\pi^{+}\pi^{-})J/\psi$~\cite{gamma3}.
Kinematting fitting of the data leads to the result,
$\bm{\mathcal{B}(J/\psi\to3\gamma)=(1.2\pm 0.3 \pm 0.2)\times10^{-5}~
(\textbf{Significance}\sim6\sigma}).$
Fig.~6 shows background subtracted data and signal Monte-Carlo
distributions. 

\begin{figure}
\includegraphics[width=2.1in]{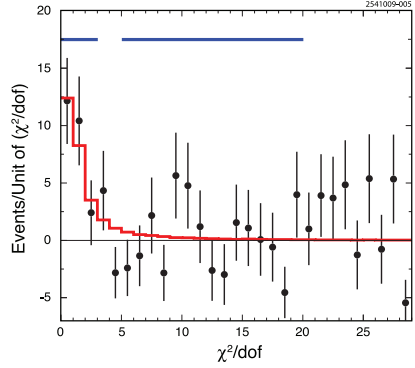}
\caption{Observation of $J/\psi\to 3\gamma$. 
Background subtracted data and signal Monte-Carlo distributions
for variable $\chi^2/dof$ of kinemattic fit. The signal and background 
normalization ragions are shown by full horizontal, 
 $\chi^2/dof=0-3$(signal), $\chi^2/dof=5-20$(background) lines.}
\end{figure}

\subsection{Precision Measurements of Branching Fractions}

 Using the data set of \textbf{26 million} $\bm{\psi(2S)}$, CLEO has
 made precision measurements of decays of $\bm{\psi(2S)}$,~$\bm{J/\psi(1S)}$,
and $\bm{\chi_{cJ}(1P)}$~\cite{all}.  Among the decays measured are:\\
\begin{tabular}{ll}
$\psi(2S), J/\psi\to \gamma h$  ($h=\pi^0,\eta,\eta'$) & \\
$\psi(2S), J/\psi\to \gamma gg$  & \\
%$\chi_{cJ}\to$ $\pi\pi,~KK,~\eta\eta,~\eta\eta',~\eta'\eta'$ & \\
$\chi_{cJ}\to {X}\overline{X}$ ($X=p,\Lambda,\Sigma,\Xi$)~~~(6 decay modes) & \\
$\chi_{cJ}\to h^+h^-,h^0h^0,~h^+h^-h^0h^0$ ($h^\pm=\pi^\pm,K^\pm,h^0=\pi^0,\eta,\eta',K^0$)~~~(12 decay modes)&  \\
$\chi_{cJ}\to\gamma\gamma$,~$\gamma V$ ($V=\rho,~\omega,~\phi)$ & \\
\end{tabular}

Many of these decays have been measured for the first time, and others have greater precision than the results in the literature.  
Some of the interesting theoretical problems that the branching fractions pose are:
\begin{itemize}
\item The ratio
$\bm{\mathcal{B}(\psi(nS)\to\gamma\eta)/\mathcal{B}(\psi(nS)\to\gamma\eta')}$
is expected to be $\sim$equal for 1S and 2S states; CLEO measured
an order of magnitude difference between the two,
(21.1$\pm$0.9)\% for 1S, and $<$1.8\% for 2S. 
\item The measured rates 
$\bm{\mathcal{B}(\chi_{c1}\to\gamma\rho)}$ and
$\bm{\mathcal{B}(\chi_{c1}\to\gamma\omega)}$ are significantly 
higher than those predicted by pQCD.
\item The ratio 
$\bm{\mathcal{B}(\chi_{c0}\to\gamma\gamma)/\mathcal{B}(\chi_{c2}\to\gamma\gamma)}$ disagrees with pQCD expectations. This result provides experimental
confirmation of the inadequacy of the present first order radiative 
corrections.
\end{itemize}

\subsection{Hadronic Decays of $\chi_{bJ}(1P,2P)$, and  Inclusive $\chi_{bJ}(1P,2P)$ Decays to Open Charm}

No hadronic decays of $\chi_{bJ}(1P)$ have been measured before.  
For $\chi_{bJ}(2P)$ the only hadronic decays measured so far 
were $\chi_{bJ}(2P)\to\pi\pi\chi_{bJ}(1P)$ and 
$\chi_{bJ}(2P)_{b1,2}\to\omega\Upsilon(1S)$.

At CLEO we have made the first measurements of 14 different decays 
of $\chi_{bJ}(1P,2P)$ to light hadrons~\cite{chibj}. Up to 12 particles were detected.
The branching fractions for the corresponding decays of $\chi_{b1,2}(1P)$ 
and $\chi_{b1,2}(2P)$ were found to be nearly equal. 
The ratios between decays to $n$ charged pions and $(n-2)$ charged +2 
neutral pions were found to approximately follow the expectations based 
on combinatorics.

CLEO also measured the inclusive decays of   
$\chi_{bJ}(nP)\to D^0+X$~\cite{d0x}.
The enhanced rates for $\chi_{b1}(1P,2P)\to D^0+X$ were found to be
consistent with NRQCD predictions.

\section{Summary}

CLEO data at  $\psi(2S)$ and $\Upsilon(1S,2S,3S)$ resonances
were analyzed. The prominent results are the following.

\begin{itemize}
\item Observation of $\bm{\eta_c'(2S)}$ in  $\gamma\gamma$ fusion, and 
its mass measurement. Search for  $\eta_c'(2S)$ in exclusive decays, 
and upper limit measurements for decay branching fractions.

\item Observation of $\bm{h_c(1P)}$ in $\psi(2S)\to \pi^{0}h_c$, 
and precision measurement of its mass.
Evidence of $h_c$ decay in multi-pion exclusive final state.

\item Confirmation of $\bm{\eta_b(1S)}$ observation, and measurement of 
mass of $\eta_b$ and decay branching fraction 
$\mathcal{B}(\Upsilon(3S)\to\gamma\eta_b)$.

\item Observation of decay $\bm{J/\psi\to 3\gamma}$ (first observation of
meson decay in $3\gamma$).

\item  Precision measurements of decay branching fractions of
  $\bm{\psi(2S)}$,~$\bm{J/\psi(1S)}$, and $\bm{\chi_{cJ}(1P)}$ 
charmonium states, and $\bm{\chi_{bJ}(1P,2P)}$ bottomonium states. 
 Many of these decays have been measured for the first time, and 
others have much greater precision than the results in the literature.

\end{itemize}

 There is a rich program of hadronic physics at CLEO; too extensive
to cover it all in one talk. There are also quite a few analyses in
a preliminary stage. Expect new results in the coming years.

%\begin{theacknowledgments}

%\end{theacknowledgments}

\bibliographystyle{aipproc}   % if natbib is available

\bibliography{sample}

\IfFileExists{\jobname.bbl}{}
 {\typeout{}
  \typeout{******************************************}
  \typeout{** Please run "bibtex \jobname" to optain}
  \typeout{** the bibliography and then re-run LaTeX}
  \typeout{** twice to fix the references!}
  \typeout{******************************************}
  \typeout{}
 }

%%%%%%%%%%%%%%%%%%%%%%%%%%%%%%%%%%%%%%%%%%%
%% The following lines show an example how to produce a bibliography
%% without the help of the BibTeX program. This could be used instead
%% of the above.
%%%%%%%%%%%%%%%%%%%%%%%%%%%%%%%%%%%%%%%%%%%

%\begin{thebibliography}{9}

%\endinput
%%
%% End of file `template-8s.tex'.

\end{document}